\documentclass{emulateapj}



\def\***#1{{\sc #1}} \def\plan#1{\relax} \def\Plan#1{\relax}
\def\PLAN#1{\relax}

\def\lta{\mathrel{\spose{\lower 3pt\hbox{$\mathchar"218$}} \raise
2.0pt\hbox{$\mathchar"13C$}}} \def\gta{\mathrel{\spose{\lower
3pt\hbox{$\mathchar"218$}} \raise 2.0pt\hbox{$\mathchar"13E$}}}
\newcommand{\etal}{{\it et al.}}  

\shortauthors{Senorita Devi et al.}
\shorttitle{Ultra-Luminous X-ray source in NGC 6946}

\def\mathnew{\mathsurround=0pt}

\def\simov#1#2{\lower .5pt\vbox{\baselineskip0pt \lineskip-.5pt
\ialign{$\mathnew#1\hfil##\hfil$\crcr#2\crcr\sim\crcr}}}

\begin{document}

\title{The spectral and temporal properties of an Ultra-Luminous X-ray source in
 NGC 6946}

\author{A. Senorita Devi\altaffilmark{1},  R. Misra\altaffilmark{2}, K. Shanthi\altaffilmark{3} and  K. Y. Singh\altaffilmark{1} }

\altaffiltext{1}{Department Of Physics, Manipur University, Canchipur,
Imphal-795003, Manipur, India; senorita@iucaa.ernet.in}

\altaffiltext{2}{Inter-University Center for Astronomy and
Astrophysics,  Post Bag 4, Ganeshkhind, Pune-411007, India;
rmisra@iucaa.ernet.in}

\altaffiltext{3}{UGC Academic Staff College, University of Mumbai, 
Mumbai-400098, India}

\begin{abstract}

 We report  variability of the X-ray source, X-7, in NGC 6946,
during a 60 ksec {\it Chandra} observation when the count rate
decreased by a factor of $\sim 1.5$ in $\sim 5000$ secs. Spectral fitting of
the high and low count rate segments of the light curve reveal that the
simplest and most probable interpretation is that the X-ray spectra are due to
disk black body emission with an absorbing hydrogen column density equal to
the Galactic value of $2.1 \times 10^{21}$ cm$^{-2}$. During the variation, the
inner disk temperature decreased from $\sim 0.29$ to $\sim 0.26$ keV  while the inner disk radius
remained constant at $\sim 6 \times 10^8$ cm. This translates into a luminosity variation 
from $3.8$ to $ 2.8 \times 10^{39}$
ergs cm$^{-2}$sec$^{-1}$ and a black hole mass of $\sim 400 M_\odot$. More complicated models
like assuming intrinsic absorption and/or the addition of a power-law component imply a higher
luminosity and a larger black hole mass. Even if the emission is beamed by a factor of $\sim 5$, the
size of the emitting region would be $> 2.7 \times 10^8$ cm  implying a black hole mass $> 180 M_\odot$. 
Thus, these spectral results provide strong evidence that the mass of the black hole in this source
 is definitely 
$> 100 M_\odot$ and more probably $\sim 400 M_\odot$.

\end{abstract}


\section{Introduction} 

{\it Chandra} observations of nearby galaxies, have confirmed the presence
of non-nuclear X-ray point sources 
\citep{Kaa01,Mat01,Zez02}, which have luminosities $> 10^{39}$ ergs/s and hence
have been called  Ultra luminous X-ray Sources (ULX).

Since
these sources radiate at a rate greater than the Eddington
luminosity for a ten-solar mass black hole, they are believed to
harbor a black hole of mass $10 \, M_\odot\! < \! M \! <\! 10^5
\,M_\odot$ \citep{Col99,Mak00}
where the upper limit 
is constrained by the argument that a
more massive black hole would have settled into the nucleus due to
dynamical friction \citep{Kaa01}. If this interpretation is true,
than these black holes have mass in the intermediate mass range between those of
stellar mass black holes found in Galactic X-ray binaries and those associated
with  Active Galactic Nuclei and hence are called Intermediate Mass Black Holes (IMBH).
For a review see \cite{Mil04} and \cite{Mil05}. 

The creation of such
black holes \citep{Por02,Tan00,Mad01} and the process by which they sustain
such high accretion rates \citep{Kin01}, are largely unknown and and when understood are expected to
make radically shifts in the present paradigms of stellar and binary evolution
and the history of the Universe.  On the other hand,  alternate models for ULX 
challenge our present understanding of accreting systems such as super-Eddington disks 
\citep{Beg02} or emission  that is 
beamed from a geometrically thick accretion disk \citep{Kin01}. For the latter case, 
it has been argued that such thick "funnel" shaped disks enhance the observed flux by just 
a factor of few \citep{Mis03}. Thus, it is important to ascertain whether ULX do indeed
harbor IMBH or not.

Since a more direct measure of the mass such as spectroscopic mass function measurement
of the binary, is not possible for ULX,  indirect evidences have to be used. One such
way is to look for similarities in the spectral and temporal properties of  ULX and black hole
X-ray binaries. There are theoretical indications that the
nature of the accretion flow should depend on the Eddington ratio $L/L_{edd} $
rather than on the actual values of the bolometric luminosity $L$ and the
Eddington limit $L_{Edd}$. Thus, a ULX should display
similar spectral and temporal characteristics as a black hole X-ray binary accreting at a
similar $L/L_{Edd}$, even though the masses of the black holes are different.

If ULX harbor IMBH, they should display analogues spectral states to X-ray binaries. Based on such a
analogy, \cite{Yua07} modeled the X-ray spectrum detected by {\it Chandra} of
the ULX X-1 in M82, within the framework of Advection Dominated Accretion Flows (ADAF)
which successfully explains the hard state spectra of X-ray binaries. They found that if
the source is likened to the low-luminosity hard state then the black hole mass
should be $M \sim 10^5 M_\odot$ else it should be $\sim 10^4 M_\odot$ if the system
 is to be compared
with the high luminosity hard state. This degeneracy occurs because
the observed spectrum during an off axis observation by {\it Chandra} ( the source is affected
by count pile-up for normal observations) was a featureless power-law as it should be if it
is analogues to the hard state. XMM observations of the source reveal a more complex
turnover at around $\sim 8$ keV \citep{Agr06}, which could either be because the 
spectral state was different during the XMM observation or that there were 
serious contamination from nearby sources. However, \cite{Sto06} report that
many ULX observed by XMM-Newton do show such high energy turnovers, which makes the
comparison with the hard state of black hole binaries ambiguous.

A more robust argument would be provided if a ULX reveals soft state like
spectral property. In the pure soft state of X-ray binaries, the
luminosity is dominated by emission from a standard disk that extends to the last
stable orbit. During this state, the high energy power-law
component (which is dominant during the low and intermediate states) contributes
$ < 5$\% of the total luminoisty. The state occurs when the Eddington ratio $> 0.02$.
 During luminosity variations (typically for X-ray novae)
the constancy of the inner most radius and it's value being always close to the predicted
last stable orbit is taken as strong evidence for the correctness of the model. In fact,
based on the model, recent attempts have been made to estimate the spin of black holes by measuring the
subtle strong gravity effects of light bending and red-shift on the spectra \citep[e.g.][]{Mcc06}.
The analogy to ULX predicts that they should exhibit a similar disk black body
component with a smaller inner disk temperature, due to the larger black hole mass.
Indeed, \cite{Mil03} report  the presence of a power-law spectrum with a cool accretion disk 
component ($kT_{in} \sim 0.1-0.5 $ keV) in
 NGC 1313 X-1 and X-2 corresponding to black hole masses $M \sim 10^{4} M_\odot$.
A comparison of ULX which show such soft components with X-ray binaries has
been undertaken by \cite{Mil04} who argue that the systematic lower temperatures
indicate that the systems harbor IMBH, but also caution for potential weakness
of such interpretations. The primary issue is the possibility of incorrect
spectral modeling especially of data with low counts \citep{Gon06}. The soft component is
strongly effected by absorption and errors in the estimation of the column density may crucially
affect the results. Another
perhaps inconsistent aspect is that for these systems, the power-law component is
nearly as luminous as the disk one and hence an analogy has to be made
with the rarer Very High  state (VHS) rather than the more frequent 
pure soft state. Moreover, since the power-law component is an important
contributer to the flux, the modeling of the disk component is suspect to
the uncertainties in modeling the power-law one.

The existence of ULX in different spectral states may also be revealed in
a systematic spectral study of a large sample of potential sources
\citep{Win06}. \cite{Swa04} fitted the spectra of such a large sample
with an absorbed power-law model and found no bimodal distribution on the
spectral index. However, using a disk black body model, a bimodal distribution 
was revealed at least for high luminosity sources for samples obtained from 
XMM-Newton \citep{Win06} and {\it Chandra} \citep{Dev07}. One set of high luminosity
sources have  temperatures $kT \sim 0.1$ keV while the other have systematically
higher temperatures $kT \sim 1 $ keV. While the higher temperature sources do
not seem to have an analogy with black hole binaries, the lower temperature one may
be identified as the equivalent of the soft state.
For a sample consisting of sources observed by {\it Chandra} which were
not affected by pile-up and which were not located in regions of excessive diffuse
emission, \cite{Dev07} fitted both an absorbed power-law and a disk black body spectral model to
ascertain the dependency of luminosity on the spectral model used. They identified a highly
luminous source, X-7, in NGC 6946 as having a soft spectrum and with 
sufficiently high counts for more detailed studies

In the next section we report on spectral and temporal analysis of the {\it Chandra}
observation of this source and summarize and discuss the results in the last section.

\section{Spectral and Temporal properties of NGC 6946, X-7}

While the temporal behavior of all the sources examined by \cite{Dev07} will be presented
elsewhere, here we report on the ksec variability observed for source, 
X-7, (R.A: 20 35  0.13  and Dec: +60 9 7.97)  in 
NGC 6946. The source is a known variable source \citep{Liu05,Lir00} and has been called 
IXO 85 \citep{Col02} and source no. 56 \citep{Hol03}.

NGC 6946 was observed by {\it Chandra} for an effective time of 59 ksec on 7 September, 2001 (Observation ID: 1043). The data reduction and analysis were done using CIAO3.2 and
HEASOFT6.0.2. Using a combination
of CIAO tools and calibration data, the source (and background) spectrum and light curve were
extracted. Spectral
fitting was done over the energy range $0.3$ to $8$ keV and spectra were rebinned such that each bin
had a minimum of 40 counts.
The distance to the source has been estimated to be between 5.1 Mpc \citep{DeV79}
and 5.9 Mpc \citep{Kar00} and hence we adopt here a distance of 5.5 Mpc.

\begin{deluxetable} {lcccc}
\tabletypesize{\scriptsize}
\tablewidth{0pt}
\tablecaption{Spectral properties of NGC 6946, X-7 }
\tablehead{
\colhead{} & \colhead{$kT_{in}$ (keV)} & \colhead{$R_{in} (\times 10^8$ cm) } & \colhead{$L (\times 10^{39}$ erg/sec) } & \colhead{$\chi^2$/d.o.f } }
\startdata

HCR & $0.29^{+0.03}_{-0.03}$ & $6.0^{+1.5}_{-1.6}$ & $3.7^{+0.5}_{-0.4}$ & $10.9/16$ \\

 LCR & $0.26^{+0.03}_{-0.03}$ & $6.0^{+1.6}_{-1.4}$ & $2.8^{+0.4}_{-0.4}$ & $22.0/14$ \\

\enddata
\tablecomments{Best fit spectral parameters for the high rate segment (HCR) and
the low rate segment (LCR). The model is an absorbed disk black body emission with the
column density fixed at the Galactic value $ 2.1 \times 10^{21}$ cm$^{-2}$.
The inner radius of the disk $R_{in}$ is 
computed from the normalization of the disk black body component using the distance to
the source $D  = 5.5$ Mpc, the viewing angle cos$i = 0.5$ and color factor $f = 1.7$. }

\end{deluxetable}

\begin{figure}
\begin{center}
{\includegraphics[width=1.0\linewidth,angle=0]{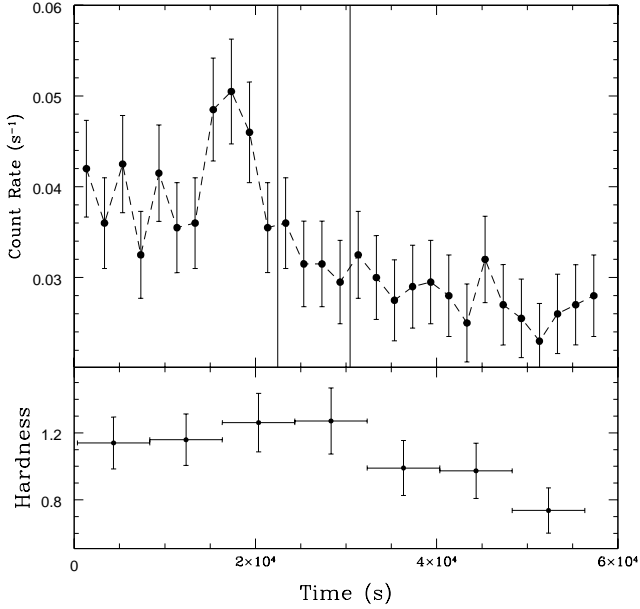}}
\end{center}  
\caption{Top panel: Light Curve of source X-7, binned over 2000 secs. A clear decrease of counts
by a factor $\sim 1.5$ is visible. The vertical lines mark the time which is used to
divide the light curve into high count rate (HCR) and low count rate (LCR) states for
spectral analysis. Bottom panel: Hardness ratio versus time binned over 8000 secs. The ratio is defined
as the ratio between flux in the $0.3-1.0$ keV band over that in the $1.0-8.0$ keV band. There is evidence that
the ratio decreases with the intensity.  }
\end{figure}

 The light curve binned over 2 ksec is shown in Figure 1, where
a clear transition from  high  to  low count rate  is visible.
The probability that the  count rate was a constant during the observations is $ < 2 \times
10^{-10}$. The bottom panel of the Figure shows the  hardness ratio versus time binned over 8000 secs. The ratio is defined
as the ratio between flux in the $0.3-1.0$ keV band over that in the $1.0-8.0$ keV band. There is evidence that
the ratio decreases with the intensity.
The two vertical lines mark the time ranges used for spectral analysis, with
the first $22.46$ ksec being called high count rate segment (HCR) state and the last $29.54$ 
ksec as the low count rate one (LCR). The total number of counts for the HCR segment was 820 while
for the LCR segment it was 725.

\begin{figure}
\begin{center}
{\includegraphics[width=1.0\linewidth,angle=0]{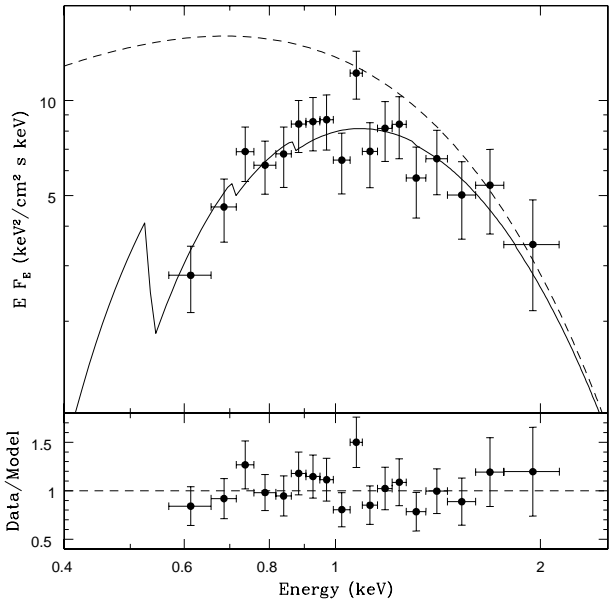}}
\end{center}  
\caption{The fitted unfolded spectra for the high count rate segment, HCR,  for parameters listed in Table 1. 
The solid line is the best fit model, while the dashed line represents unabsorbed spectra.
}
\end{figure}

\begin{figure}
\begin{center}
{\includegraphics[width=1.0\linewidth,angle=0]{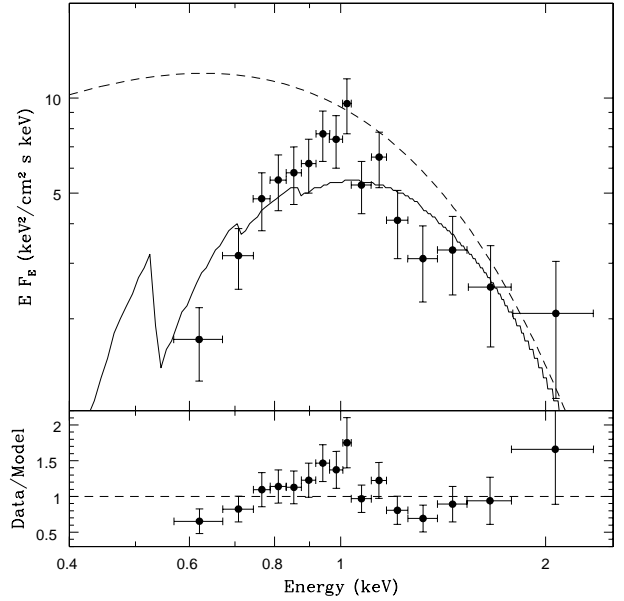}}
\end{center}  
\caption{The fitted unfolded spectra for 
the low count rate segment, LCR, for parameters listed in Table 1. 
The solid line is the best fit model, while the dashed line represents unabsorbed spectra.
}
\end{figure}

The Galactic hydrogen column density along the direction of the source is
$2.1 \times 10^{21}$ cm$^{-2}$. 
An absorbed power-law model with this value of absorption gives an unacceptable fit
to both HCR and LCR segments, with reduced $\chi^2_{\nu} > 2.5$. The fit can be significantly improved by 
allowing for intrinsic absorption, $ N_H \sim 10^{22}$ cm$^{-2}$ but with a large unphysical photon
index of $\Gamma >  5$. On the other hand, as shown in Table 1 and illustrated in Figures 2 and 3, 
an absorbed disk black body
model with the column density at the Galactic value, provides adequate fits to both segments. 
The inner radius of the disk $R_{in}$ is 
computed from the normalization of the disk black body component using the distance to
the source $D  = 5.5$ Mpc, the viewing angle cos $i = 0.5$ and color factor $f = 1.7$.
The result and it implications are better represented in Figure 4, where confidence region ellipses
are shown for the two parameters $R_{in}$ and inner disk temperature $T_{in}$. The Figure
reveals that the simplest and more favored interpretation is that the inner disk radius remains
constant, while the temperature decreases during the variation. Assuming that $R_{in} \sim 10 GM/c^2$,
the black hole mass, $M$, can be estimated and confidence ellipses for $M$ and total bolometric
luminosity are shown in Figure 5. A black hole mass of $M \sim 400 M_\odot$ is favored for
both sets while the luminosity decreases from $\sim 3.7$ to $\sim 2.8 \times 10^{39}$ ergs/sec,
corresponding to an Eddington ratio, $L/L_{Edd} \sim 0.06$.

\begin{figure}
\begin{center}
{\includegraphics[width=1.0\linewidth,angle=0]{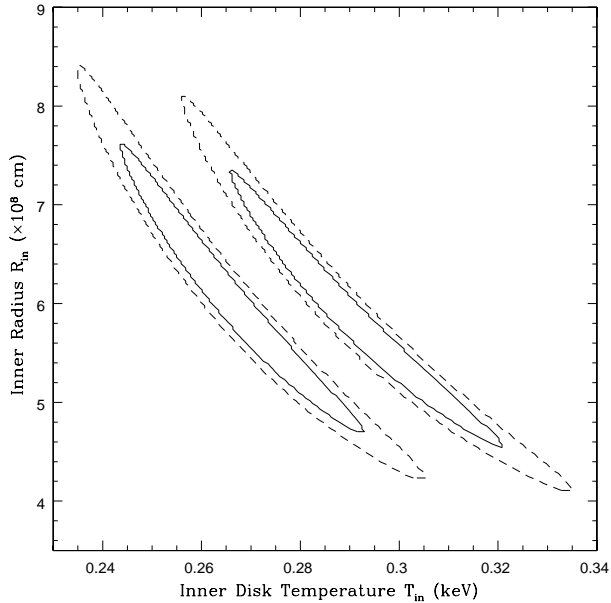}}
\end{center}  
\caption{Confidence region ellipses for the inner disk radius, $R_{in}$ and temperature, $T_{in}$, 
for the HCR and LCR segments for best fit parameters given in Table 1. The solid
(dashed) curves represent $\Delta \chi^2 = 2.3 (4.61)$ corresponding to $68.3$ ($90$\%) confidence
level. The results favor the interpretation that $R_{in}$ is a constant and that the decrease in
count rate is due to a decrease in $T_{in}$. }
\end{figure}

\begin{figure}
\begin{center}
{\includegraphics[width=1.0\linewidth,angle=0]{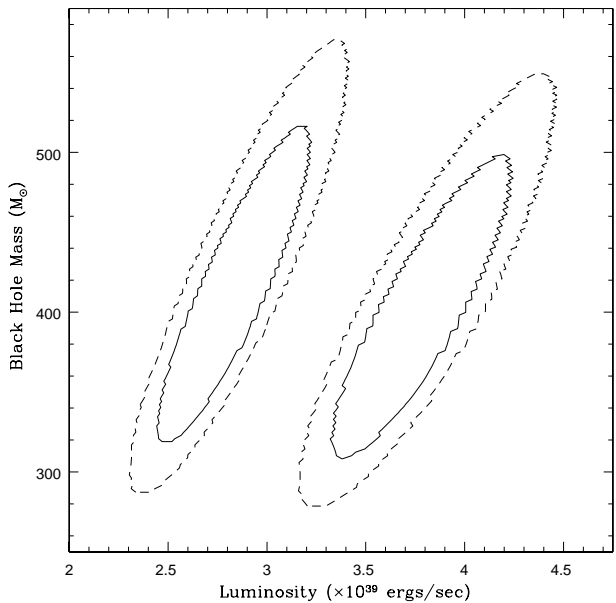}}
\end{center}  
\caption{Confidence region ellipses for the black hole mass, $M$ and bolometric luminosity, $L$, 
for the HCR and LCR segments estimated from $R_{in}$ and $T_{in}$ values shown in Figure 3. The solid
(dashed) curves represent $\Delta \chi^2 = 2.3 (4.61)$ corresponding to $68.3$ ($90$\%) confidence
levels. For both the segments a black hole mass, $M \sim 400 M_\odot$ is favored.}
\end{figure}

It is worth emphasizing that the above interpretation is the simplest one which
provides acceptable fit to the data and requires the smallest black hole mass. 
If one allows for intrinsic absorbing column density, the best fit temperature decreases
and the inner radius increases implying a larger black hole mass. For example, a 
simple model could be that the variation is only due to changes in the
intrinsic column density. A model where the column density is allowed to vary, with
$R_{in}$ and $T_{in}$ same for both segments, indeed gives an acceptable combined $\chi^2$/dof = $36.9/32$,
with $N_H$ increasing from  $3\pm 0.4$ to $3.9\pm 0.4 \times 10^{21}$ cm$^{-2}$ and inner
disk temperature $T_{in} = 0.22 \pm 0.1$ keV. The inner disk radius turns out to be
$1.5 \pm 0.3 \times 10^9$ cm corresponding to a $\sim 950 M_\odot$ black hole. If
the column density and temperature are allowed to vary independently
for the two segments, $\chi^2/dof = 29/29$, with $R_{in} = 1.3 \pm 0.4 \times 10^9$ cm
corresponding to $M \sim 900 M_\odot$. An additional power-law component with photon index
$\Gamma = 2.0$ or $\Gamma = 2.7$ does not improve the fit if the column density is fixed
at the Galactic value. For $N_H \sim 6 \times 10^{21}$, a disk black body and a power-law component
with $\Gamma = 2$, provides a good fit to both segments $\chi^2/dof = 19.9/27$. Here, the variability
is due to a change of normalization of the power-law component, while the temperature
$T_{in} \sim 0.13$ keV and $R_{in} \sim 1.2 \times 10^{10}$ cm remain  nearly constant. Again, in
these more complicated interpretations $R_{in}$ is significantly larger implying a larger black hole
mass.

\section{Summary and Discussion}

The X-ray source X-7 in NGC 6946 is located in a region of low diffuse emission and has
a count rate which does not cause pile-up in  {\it Chandra} detectors. The light curve
of this source reveals a decrease of $\sim 1.5$ in the count rate over $5000$ sec making it 
one of the few ULX that have clearly shown variability on ksec time-scales \citep{Kra05,Miz01}.

Spectral fitting of the high and low count rate parts of the light curve reveals that
a  simple model of a disk black body emission absorbed by the Galactic column density of
$N_H = 2.1 \times 10^{21}$ cm$^{-2}$ can adequately represent both segments. The best fit temperature
varies from $0.26$ to $0.29$ keV while the inner disk radius remains constant at $ \sim 6 \times 10^8$ cm.
This would imply that the mass of the black hole is $\sim 400 M_\odot$ and at a luminosity of
$\sim 3 \times 10^{39}$ ergs/sec the system has an Eddington ratio of $\sim 0.06$. Other more
complicated spectral fits like assuming intrinsic column density for the source and/or addition
of a power-law component results in a larger estimation of inner disk radius and hence a 
larger black hole mass. Although the luminosity of the source estimated here is only
mildly super-Eddington for ten solar mass black hole, the low inner disk temperature
$\sim 0.3$ keV implies a larger inner disk radius and hence a large black hole mass $\sim 400 M_\odot$.
A low sub-Eddington accretion rate on a ten solar mass black hole could produce the observed low inner disk temperature, but then
the predicted luminosity would be significantly less than what is observed.   
As cautioned by \cite{Mil04}, the source could be a background AGN and
the soft component observed is actually the soft excess which is detected in many AGN.
These soft excesses can be  modeled as black body emission at $kT \sim 0.1$ keV similar
to the soft component in ULX.
While this may be true for some ULX, this is unlikely in this case because here (unlike soft excess
of AGN) the soft component totally dominates the luminosity.

If the emission is beamed \citep{Kin01} by a factor $\eta$, the inner disk radius would be overestimated
by $\eta^{1/2}$.  However, even for extreme geometries \cite{Mis03} have computed $\eta < 5$. Hence
an extreme beaming of $\eta \sim 5$, would imply that $R_{in} \sim 2.7 \times 10^8$ cm corresponding
to a black hole mass of $180 M_\odot$. The color factor used in this analysis is the standard
value of $1.7$. Even for the extreme case that there is no color correction, 
$R_{in} \sim 2 \times 10^8$ cm corresponding to a black hole mass of $130 M_\odot$. Thus only
in the extremely unlikely case of having no color correction and high beaming factor
$\eta \sim 5$ would the black hole mass be $< 100 M_\odot$.

A model can be envisioned where the inner disk radius is not at $10 GM/c^2$ but is
at a much larger radius say $100 GM/c^2$, with the inner region being highly radiatively
inefficient. Then the mass of the black hole could be $\sim 40 M_\odot$. However to produce a 
luminosity of $\sim 2 \times 10^{39}$ ergs/sec at $100 GM/c^2$ would require an accretion rate 
which is $50$ times the Eddington value and it is not clear how the inner region would remain 
radiatively inefficient at such accretion rates. A more radical model would be that the emission
arises from an optically thick sphere of radius $ \sim 100 GM/c^2$ surrounding a $40 M_\odot$ black hole,
but that would be a serious paradigm shift from our present understanding of accretion flows. 

In a contemporary work, \cite{Fri08}, have studied the long term variability of
sources in NGC 6946 using five {\it Chandra} observations. They confirm
the variability of this source (their source no. 60) for this
observation, but not for the others. They 
detect long term flux variability for the different observations. 
Their hardness 
ratio estimations confirm the soft nature of this source for all the 
observations.

The results presented here indicate that the black hole mass of this source is $> 400 M_\odot$
which may be true for other super-soft X-ray sources and for ULX in general. However, 
further analysis of data of this and other super-soft sources are required to
ascertain whether such temporal and spectral behavior is common among ULX.

\acknowledgements

ASD thanks CSIR and IUCAA for support. The authors would like to thank
V. Agrawal, S. Ashtamkar, M. Patil and V.  Wadwalkar for helping in the
data analysis and Phil Charles for useful comments on an earlier draft.

\end{document}